\documentstyle[tm6conf]{article}

\begin{document}
\ \
\vskip 1.1cm

\title{PLANETARY NEBULA ABUNDANCES, STELLAR YIELDS, AND THE GALACTIC EVOLUTION OF $^{12}$C AND $^{14}$N}

\author{R.B.C. Henry\altaffilmark{1}, K.B. Kwitter\altaffilmark{2}, and J. Buell\altaffilmark{1}}
\altaffiltext{1}{Department of Physics \& Astronomy, University of Oklahoma}
\altaffiltext{2}{Department of Astronomy, Williams College}

\begin{resumen}
\baselineskip 14pt % DO NOT REMOVE THIS LINE
A project which aims to understand the abundance patterns in He,
C, N, O, Ne, S, and Ar for
a small sample of planetary nebulae is described.  Abundance
ratios of O/H, C/O, and N/O especially
show a broad range relative to their solar values 
and AGB models used to interpret the patterns
indicate that progenitor ranges in mass and metallicity are
adequate to explain the observed abundance ranges.  Chemical
yields of intermediate-mass stars are inferred from these same
models and are used to calculate chemical evolution models of
the solar neighborhood.  These models indicate that
intermediate-mass stars account for roughly half of the $^{12}$C
and all of the $^{14}$N in that region.
 \end{resumen}

\begin{abstract}
A project which aims to understand the abundance patterns of He,
C, N, O, Ne, S, and Ar for
a small sample of planetary nebulae is described.  Abundance
ratios of O/H, C/O, and N/O especially
show a broad range relative to their solar values, 
and AGB stellar evolution models used to interpret the patterns
indicate that reasonable ranges in progenitor mass and metallicity are
adequate to explain the observed abundance spread.  Chemical
yields of intermediate-mass stars inferred from these same
models are used to calculate chemical evolution models of
the solar neighborhood.  These models indicate that
intermediate-mass stars account for roughly half of the $^{12}$C
and almost all of the $^{14}$N in that region.
\end{abstract}
\bigskip

\keywords{\bf ISM: ABUNDANCES --- PLANETARY NEBULAE: GENERAL --- STARS: EVOLUTION
--- GALAXY: ABUNDANCES}

\section{Introduction} Intermediate-mass stars
(0.8$<$M$<$8.0M$_{\odot}$; hereafter IMS) are thought to contribute
significantly to the production of carbon and nitrogen in the Galaxy
(Tinsley 1978; Sarmienta \& Peimbert 1985).  After stars within this mass
range leave the main sequence, they pass through one or more stages
when material synthesized during core and shell-burning is mixed into
the outer envelope and subsequently expelled as a planetary nebula (PN)
is formed.  Numerous PN abundance studies indicate that
mass fractions of He, C, and N are often enhanced relative to solar and interstellar values in these objects (Henry 1990; Perinotto 1991).
The frequency of PN formation further implies that IMS may contribute
significantly to the buildup of C and N in the Galaxy.  Currently, we
are attempting to quantify this picture by coupling detailed abundance
studies of planetary nebulae with stellar and chemical evolution
models.

Our PN abundance study focuses on a small number of carefully chosen
galactic objects for which spectral information is available between
1150-9600{\AA}.  Abundances of He, C, N, O, Ne, S, and Ar are
determined through a detailed analysis.  The aim of the stellar
evolution study is to interpret PN abundance patterns by calculating models of thermally pulsating asymptotic giant branch (AGB) stars as well as the chemical composition of their ejecta as they produce PNe.
In the process, yields
(production rates) of elements such as C and N are inferred from the
models.  These yields in turn are employed in chemical evolution models
to understand the sources of carbon and nitrogen in the solar
neighborhood.  In this paper, we provide an update of the project along with some preliminary results.  Sections 2, 3, and 4 describe the three phases of the project, while section 5 is a summary.

\section{Planetary Nebula Abundance Study}

We have chosen for our study 22 PNe to span wide progenitor mass and metallicity
ranges.  Inferring abundances of the elements listed above requires that we have spectral coverage both in the UV and optical.  UV spectral data are taken from the IUE Final Archive, a
repository for data which have recently been re-reduced using
standardized reduction algorithms that have resulted in improved
spectral resolution and S/N.  We then combine the UV data with
optical spectra available in the literature or (lately) spectra we have
acquired ourselves.  We are currently working to extend spectral
coverage for all of our objects out to around 9600{\AA}.  When
completed, we will have spectra available from 1150 to 9600{\AA} for
each of our objects.

Prior to performing an abundance analysis, we correct the spectra
for effects of interstellar reddening and then merge the UV and optical
data using line ratios comprised of lines occuring in each of the two
spectral regions and whose values can be predicted from theory.  Two
ratios we have found useful for this purpose are He~II
$\lambda$1640/$\lambda$4686 and C~III]
$\lambda$1909/C~II $\lambda$4267.

The abundance of an element $X$ is then determined using the relation:

$$X = icf(x) \times \xi(X) \times {\sum^{obs}}{{I_{\lambda}}\over{\epsilon_{\lambda}(T_e,N_e)}},$$

\noindent where $I_{\lambda}$ is the measured intensity of a spectral feature
produced by an ion of element $X$, $\epsilon_{\lambda} (T_e,N_e)$ is the energy production rate per ion
of the spectral feature $\lambda$.  Quotients for all observable ions are summed together.  Then,  $icf(X)$ is the ionization correction
factor, i.e. the ratio of the sum of abundances of {\it all} $X$ ions
to the sum of {\it observed} ions of $X$, and

$$\xi(X) = {{true~model~abundance}\over{apparent~model~abundance}}.$$

The factor $\xi(X)$ is determined by first calculating a detailed
photoionization model to reproduce the observed values of numerous diagnostic
line ratios which are sensitive to physical conditions along the
line-of-sight, and then applying a modified version of the above
abundance method to the output spectrum of a photoionization model to
determine an ``apparent'' abundance.  Then, by comparing this
quantity with the ``true'' model input value for the same element, $\xi(X)$ is
obtained.

Results for the 16 objects analyzed thus far are presented in Fig.~1,
where we plot values of He/H, O/H, C/O, N/O, and Ne/O, all normalized to the sun (linear scale for He/H;
logarithmic scale for all others).  Work on S and Ar abundances is
currently in progress, and results for these are not included here.  Abundance ratios for sample PNe are indicated with open circles.  Local interstellar values from Snow \& Witt (1996) are shown with plusses.
Because of the difficulty of showing detailed
abundance information for individual objects in a plot such as this, we
refer the reader to our papers where these results are presented in
more detail: Henry, Kwitter, \& Howard (1996), Kwitter \& Henry (1997),
and Kwitter \& Henry (1998).  What we wish to point out here is the
{\it broad range} in abundances which is apparent among the 16 PNe.  For example, C/O and N/O for our sample span roughly 3 and 2 orders of magnitude, respectively.  We want to emphasize that this scatter is for the most part real and cannot be explained by abundance uncertainties from the procedure itself, which are roughly 0.3~dex.  Understanding this variation in abundance is one of the principal goals of the stellar evolution models described in the next section.

\section{AGB Models}

Intermediate-mass stars go through as many as three important stages in
their post main sequence evolution which result in synthesized material from
their interiors being brought to the surface and eventually expelled by wind or
during PN formation.  These stages are:

\begin{itemize}

\item 1st dredge-up, which occurs during the red giant phase when convection
reaches into the H-exhausted core, bringing up $^4$He and $^{14}$N.

\item 2nd dredge-up, which occurs in stars of M$\ge$5M$_{\odot}$, early in
the AGB phase, when additional $^4$He and $^{14}$N are brought up.

\item 3rd dredge-up, which occurs during the thermally pulsing AGB phase when
convection associated with the interpulse phases reaches into the
He-burning shell, bringing up triple-alpha products.  Subsequent
burning can also occur at the base of the convective envelope when
$^{12}$C is converted to $^{14}$N and $^{22}$Ne, a process known as hot bottom burning.

\end{itemize}

Thermally pulsing AGB models have been computed by J.~Buell as part of
his PhD thesis research at the University of Oklahoma.  Buell has
acquired the code used in Renzini \& Voli (1981) and made significant
updates, including the introduction of an \.M-period relation and the
opacities of Rogers \& Iglesias (1992).  The code assumes a set of envelope
abundances from first and second dredge-ups and proceeds to calculate a model of a
thermally pulsing AGB envelope which includes the effects of 3rd
dredge-up and hot bottom burning.  The calculation follows the
evolution of $^4$He, $^{12}$C, $^{13}$C, $^{14}$N, and $^{16}$O during
each interpulse, eventually predicting the masses of these isotopes
which are lost due to winds or ejected during PN formation.  The predicted PN abundances are then compared with
abundances inferred from observations, as described in the previous
section.

Results of Buell's study are presented in Buell (1997).  Briefly, the
models explain the large range in abundances shown in Fig.~1 as being
due to ranges in progenitor mass and metallicity.  Buell's models
also predict a set of chemical yields which are used in the next
section to estimate the contribution of IMS to the chemical evolution
of $^{12}$C and $^{14}$N in the solar neighborhood.

\section{Chemical Evolution of $^{12}$C and $^{14}$N}

We have calculated evolution models for $^{12}$C and $^{14}$N for the
solar neighborhood following prescriptions described in detail by
Timmes et al. (1995).  Our models take account of production time delays
due to the significant main sequence life times of IMS, i.e.
instantaneous recycling is {\it not} assumed.  In addition, we assume a
star formation rate (SFR) such that $SFR \propto M_T
\left({{M_g}\over{M_T}}\right)^2,$ where $M_T$ and $M_g$ are total and
gas mass, respectively.  An infall rate was assumed to be exponentially
decreasing with time, with a characteristic time scale of 4~Gyr.  The age
of the Galaxy was taken to be 15~Gyr.  IMS yields were taken from Buell
(1997) while yields for massive stars (M$\ge$13M$_{\odot}$) and Type~Ia supernovae were taken from Nomoto (1997a,b).

The chemical evolution model results for $^{12}$C and $^{14}$N are
presented in Fig.~2, where the mass ratios of C/H and N/H are plotted
versus time in Gyr.  Total mass ratios as well as contributions from massive stars only are represented with bold and dashed lines, respectively.
The model results indicate that intermediate-mass and massive stars contribute
equally to the buildup of $^{12}$C.  However, IMS are apparently
responsible for most of the $^{14}$N in the solar neighborhood.  These results will be presented in greater detail in a future paper.

\section{Summary}

Careful abundance analysis of a small sample of planetary nebulae
reveals large and real scatter in abundance ratios He/H, O/H, C/O, and
N/O.  Using detailed AGB models calculated with a newly-updated code,
this scatter is in turn explained by broad ranges in mass and
metallicity of the progenitor stars.  Carbon and nitrogen yields
inferred from these same models are employed in chemical evolution
models to show that IMS produce most of the nitrogen and roughly half
of the carbon in the solar neighborhood.

\begin{figure}
\figurenum{1}
%\epsscale{.8}
\plotone{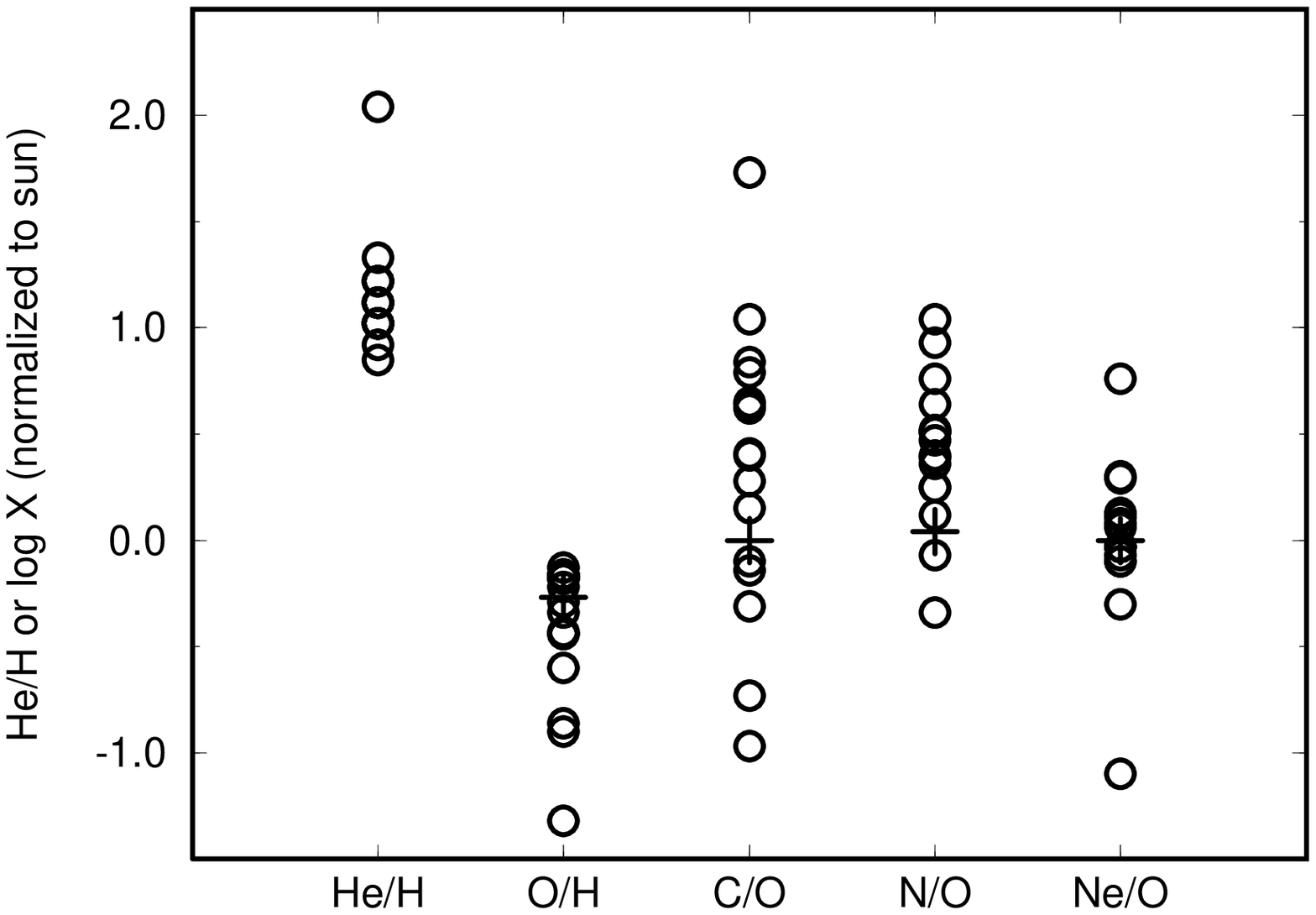}
%\plotfiddle{fig1.ps}{}{0}{80}{80}{0}{72}
\caption{Abundance ratios for 16 PNe normalized to the sun.  Values for He/H are on a linear scale, while those for the remaining ratios are on a logarithmic scale.  PN abundances are indicated with open circles.  The plus symbol shows the local interstellar values as determined by Snow and Witt (1996)}
\figurenum{2}
%\epsscale{1}
\plottwo{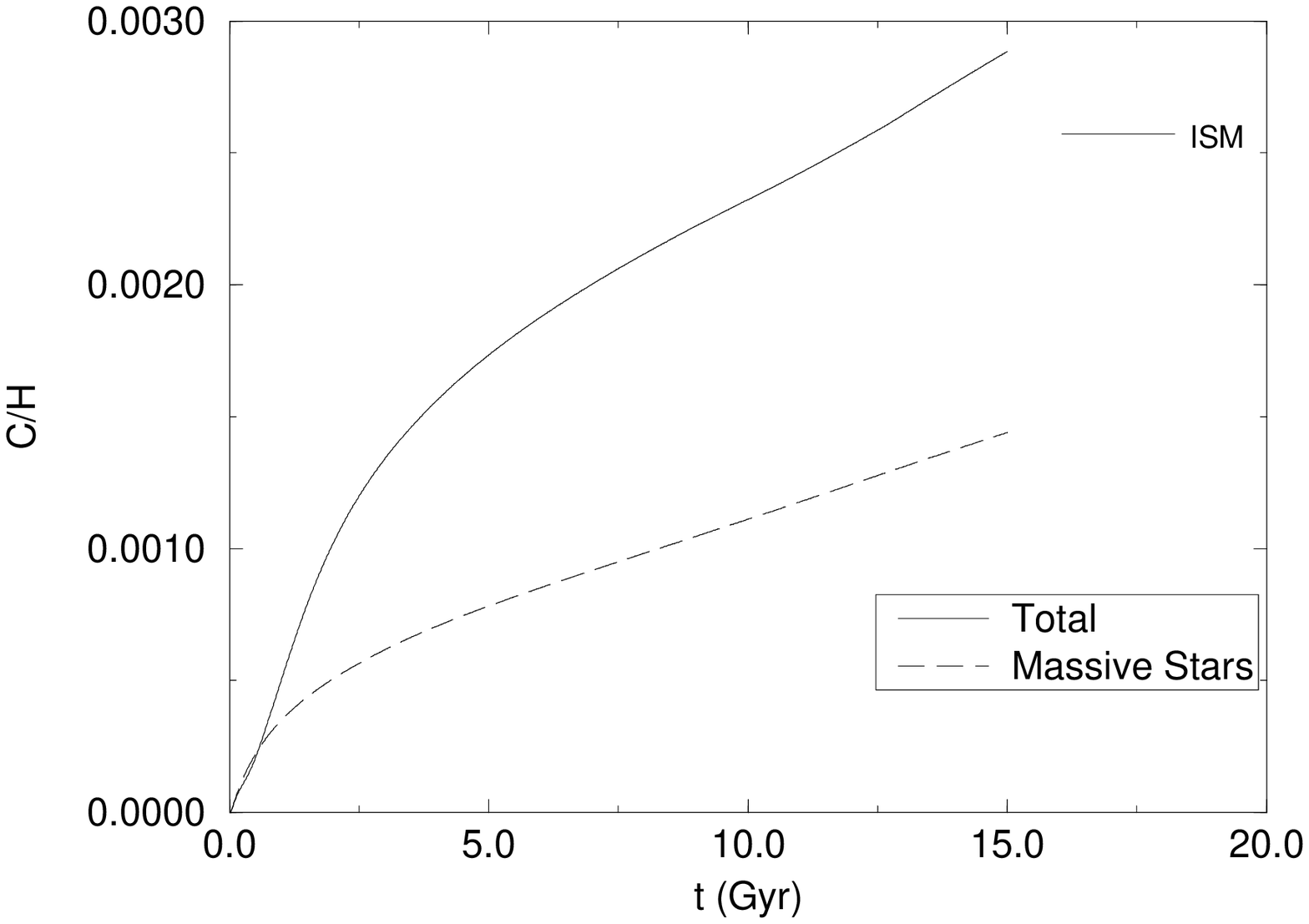}{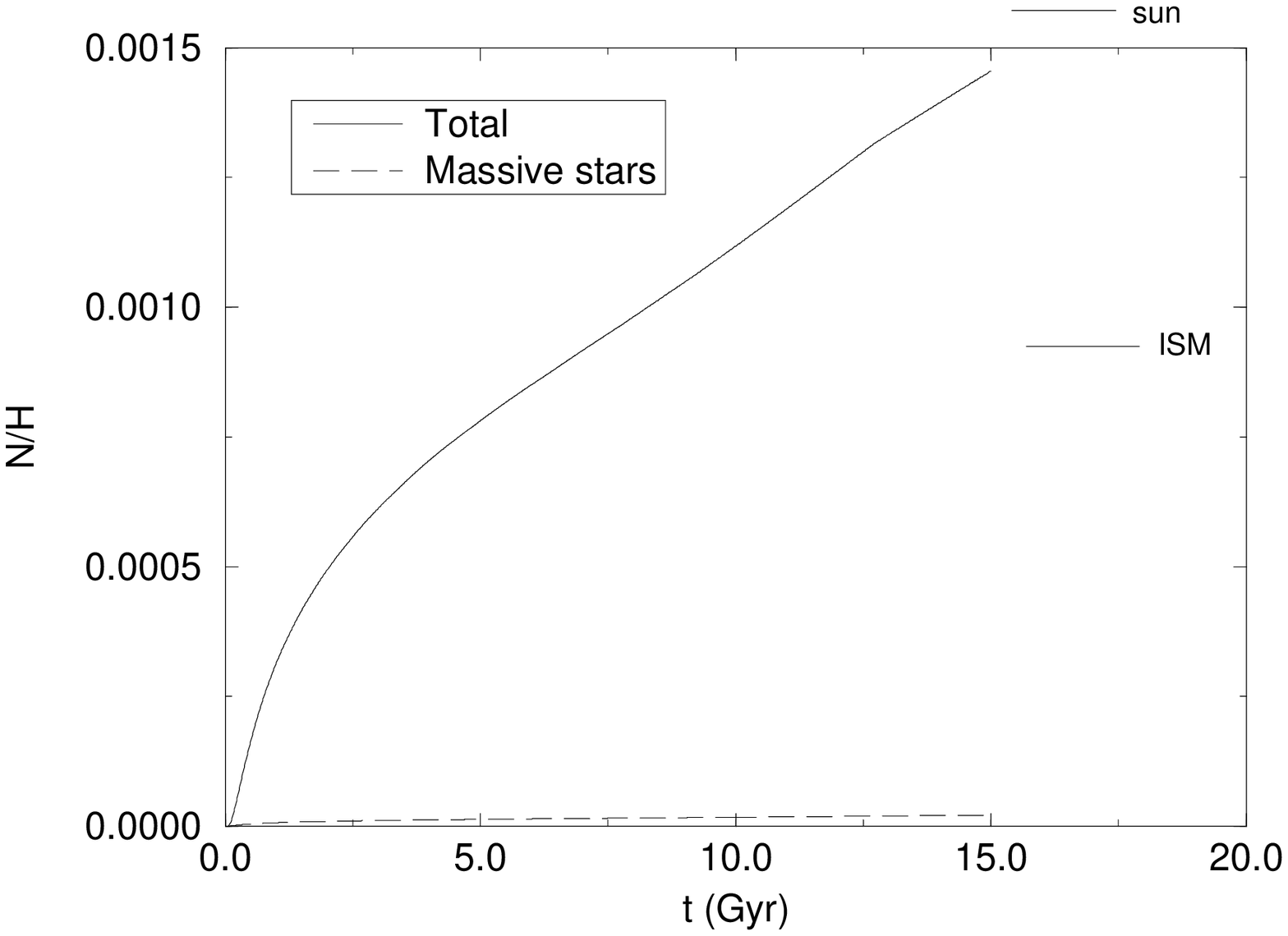}
\caption{a. Predicted mass ratio of C/H versus time in Gyr for the solar neighborhood.   b. Same as a. but for N/H.  Local interstellar (Snow \& Witt 1996) and solar (Anders \& Grevesse 1989) values are shown.}
\end{figure}

\end{document}